\documentclass[twocolumn,aps,prl,showpacs]{revtex4}

\usepackage{epsfig}

\begin{document}

\title{Fracture precursors in disordered systems}

\author{ G. Meacci$^{1,2}$, A. Politi$^{1}$, and M. Zei$^{3}$}

\affiliation{
$^{1}$
Istituto Nazionale di Ottica Applicata, L.go E. Fermi 6, 50125 Firenze, Italy\\
$^2$ Max Planck Institute for Physics of Complex Systems, N\"othnitzer Stra\ss e
38, 01187 Dresden, Germany\\
$^3$
CECM-CNRS-Vitry, 15 rue Georges Urbain, 94407 Vitry-sur-Seine, France
}



\date{\today}

\begin{abstract}
A two-dimensional lattice model with bond disorder is used to investigate the
fracture behaviour under stress-controlled conditions. Although the cumulative
energy of precursors does not diverge at the critical point, its derivative
with respect to the control parameter (reduced stress) exhibits a singular
behaviour. Our results are nevertheless compatible with previous experimental
findings, if one restricts the comparison to the (limited) range accessible 
in the experiment. A power-law avalanche distribution is also found with an
exponent close to the experimental values. 
\end{abstract} 
\vspace{2mm}

\maketitle


Fractures are very complex phenomena which involve a wide range of spatial and
sometimes temporal scales. Accordingly, the development of a general theory 
is quite an ambitious goal, since it is not even clear whether a continuous
coarse-grained description makes sense; additionally, for the very same reason,
realistic simulations are almost unfeasible. However, such difficulties
have not prevented making progress on several aspects of fracture
dynamics such as propagation velocity, roughness, or the failure time under
a constant stress \cite{FM99,B97,P92}. In this paper we are interested in
studying the development of the so-called precursors, microcracks preceding the
macroscopic fracture in a brittle disordered environment. Some recent
experiments \cite{GGBC97,GCGZS02,STA02} suggest that we are in the presence of
a critical phenomenon, although the accuracy is not yet high enough not only to
discuss its universality properties, but also to assess the order of the
transition.

Fractures are typically studied either by increasing the strain or the stress
and recording the acoustic emissions generated by the microfractures preceding
the final breakup. In both cases, there is experimental evidence that the
probability density $N(\varepsilon)$ of microfractures with energy between
$\varepsilon$ and $\varepsilon + d\varepsilon$, follows a power law
\begin{equation}
N(\varepsilon) \sim \varepsilon^{-\beta} ,
\end {equation}
although it has been found that different materials are characterized by
different values of the exponent $\beta$: $1.3$ in synthetic plaster
\cite{PPVAC94}, $1.51$ in wood \cite{GGBC97}, $1.9$ in fiberglass 
\cite{GCGZS02}, and $1.25$ in paper \cite{STA02}.

On the other hand, the cumulative energy emitted while approaching the
fracture exhibits a qualitatively different behaviour depending whether the
strain or the stress is controlled. In the former case, after a bond breaks,
lattice rearrangements lead to a stress reduction which, in turn, increases the
overall stability. As a consequence, a critical behaviour can be realistically
observed only in the latter case. For instance, in Refs.~\cite{GGBC97,GCGZS02}
it was found that
\begin{equation}
\label{energy}
E(P_r) \sim \left( \frac{P_c-P}{P_c} \right)^{-\lambda} := P_r^{-\lambda} \quad ,
\end {equation}
where $E$ is the cumulative energy released up to pressure $P$, $P_{c}$
is the critical pressure corresponding to the macroscopic failure, and $P_r$ is
the so-called reduced parameter.
\begin{figure}[tcb]
\centering
\includegraphics*[width=8 truecm, angle=0]{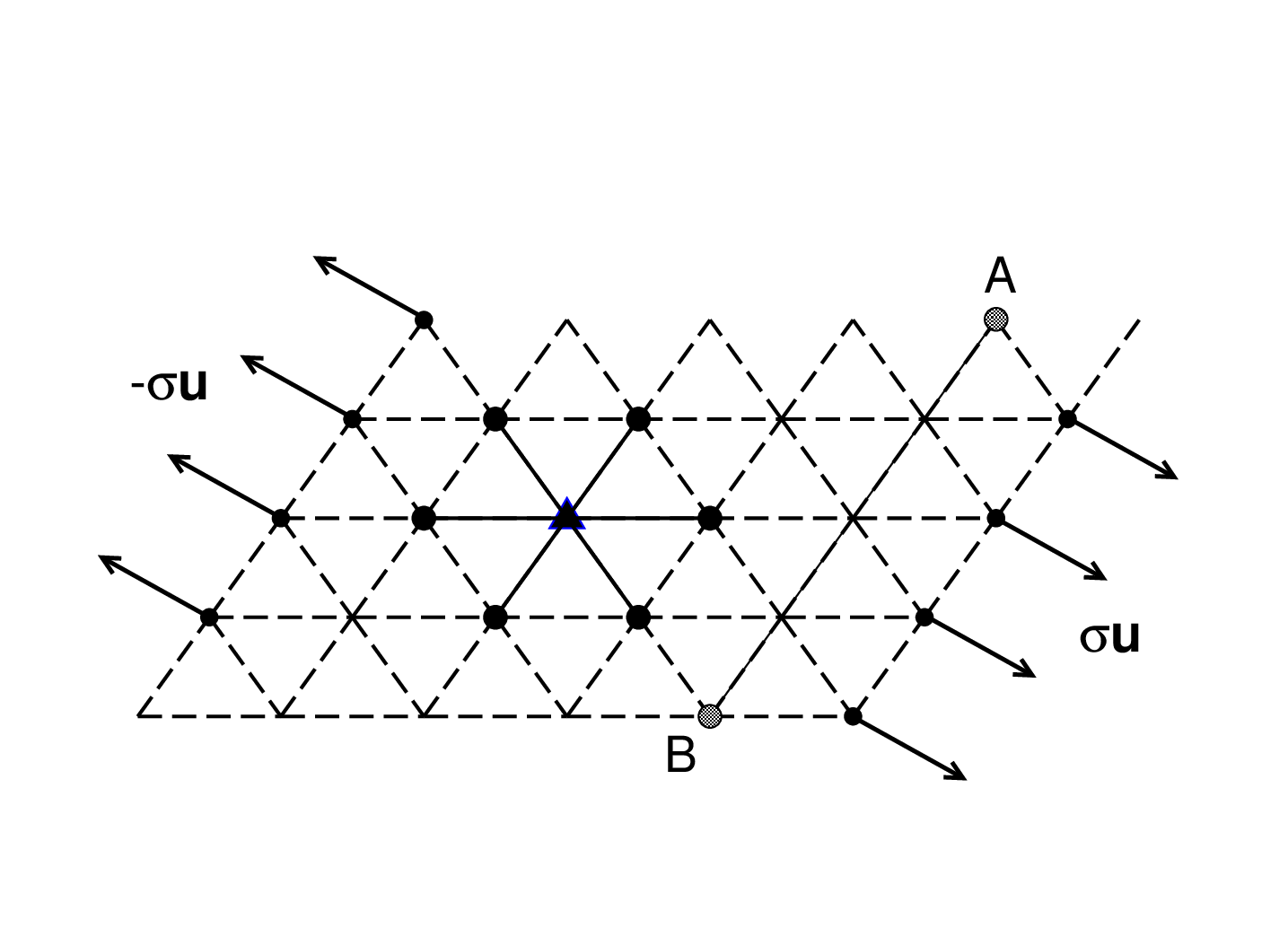}
\caption{Sketch of the triangular lattice. The solid lines refer to the nearest
neighbour interactions involving the site denoted by a full triangle. Left and
right boundaries are pulled apart by a force $\pm \sigma {\bf u}$.
Periodic boundary conditions are assumed along upper and lower borders
(e.g., the point $A$ can be identified with $B$).}
\label{lattice}
\end{figure}          
In the past, simplified models have been introduced in the hope to capture the
essential ingredients of the process. Among them, random fuse networks have
been quite popular, since only one scalar variable (the electric current) is
needed to describe the onset of a macroscopic failure\cite{HR90}. After the
concept of self-organized criticality (SOC) was introduced by
P.~Bak \cite{B96}, the possible interpretation of fractures as instances of
such critical phenomena became an appealing perspective to many researchers. As
a result, probabilitstic models inspired by the simplest SOC ideas have been
introduced and numerically investigated \cite{CDP96}. Altogether, it has
emerged a scenario of the fracture as a critical phenomenon, but the relative
``distance'' between models and physical reality leaves doubts about the
validity of such conclusions.

More realistic models where each particle feels the force field induced by its
nearest neighbours were already introduced in the
'80s\cite{CCD86,BS88,HRH89,HS89}, when preliminary studies of small 
2-dimensional
lattices with quenched disorder have been carried out. However, only in the
late '90s it has become possible to simulate sufficiently large systems to
attempt a scaling analysis. In Ref.~\cite{ZRSV99} a square lattice with two- and
three-body interactions was studied, finding $\beta = 2.5$ in agreement with the
mean-field behaviour of the fiber bundle model \cite{HH92,KHH97}, but larger
than the experimental values. On the other hand, large statistical uncertainties
prevent drawing any conclusion about the scaling behaviour of $E(P_r)$.

In this Letter, studying a slightly simpler model \cite{PZ01}, we are able to
investigate the behaviour of $E(P_r)$ obtaining results that are compatible with
the Lyon experiment \cite{GGBC97}. However, having here investigated a wider
range of parameter values, we are led to exclude a divergence of $E(P_r)$ near
the critical point.

Our model consists of point-like particles sitting on a 2d triangular lattice
with nearest-neigbour interactions mediated by central forces (see solid lines
in Fig~\ref{lattice}). More precisely, the force ${\bf f}_{ij}$ acting on
the $i$th particle due to the interaction with the $j$th one is
\begin{equation}
\label{force}
{\bf f}_{ij} = F_{ij}(|{\bf r}_i-{\bf r}_j|)
\frac{{\bf r}_i-{\bf r}_j} {|{\bf r}_i-{\bf r}_j|} \quad ,
\end{equation}
where ${\bf r}_i$ denotes the $i$th particle position, $|\cdot|$
represents the modulus operation. Moreover, $F_{ij}(x=a) =0$, with $a$ being
the lattice spacing at equilibrium and $k_{ij} = -dF_{ij}/dx |_{a} >0$ so as to
ensure stability. Moreover, as soon as the mutual distance
$| {\bf r}_i - {\bf r}_j |$ becomes larger than some threshold
$a^*_{ij}$, the interaction strength is irreversibly set equal to 0.
As the $a^*_{ij}$'s are generally chosen close to $a$ 
(see also Ref.~\cite{BS88}), only small deviations from equilibrium can be 
expected and, accordingly, the force can be linearized. Here, we have however 
preferred the equivalent choice (at the first order level)
$F(x) = -(x^{2}-a^{2})/(2a)$, since it avoids the computation of a square root
for each bond, thus ensuring faster simulations. 

Furthermore, an external force $\sigma {\bf u}$ ($-\sigma {\bf u}$) is
applied to each particle of the right (left) boundary along a direction
${\bf u}$ orthogonal to the edges, while periodic boundary conditions are
assumed along the upper and lower boundaries.
\begin{figure}[tcb]
\centering
\includegraphics*[width=8 truecm, angle=0]{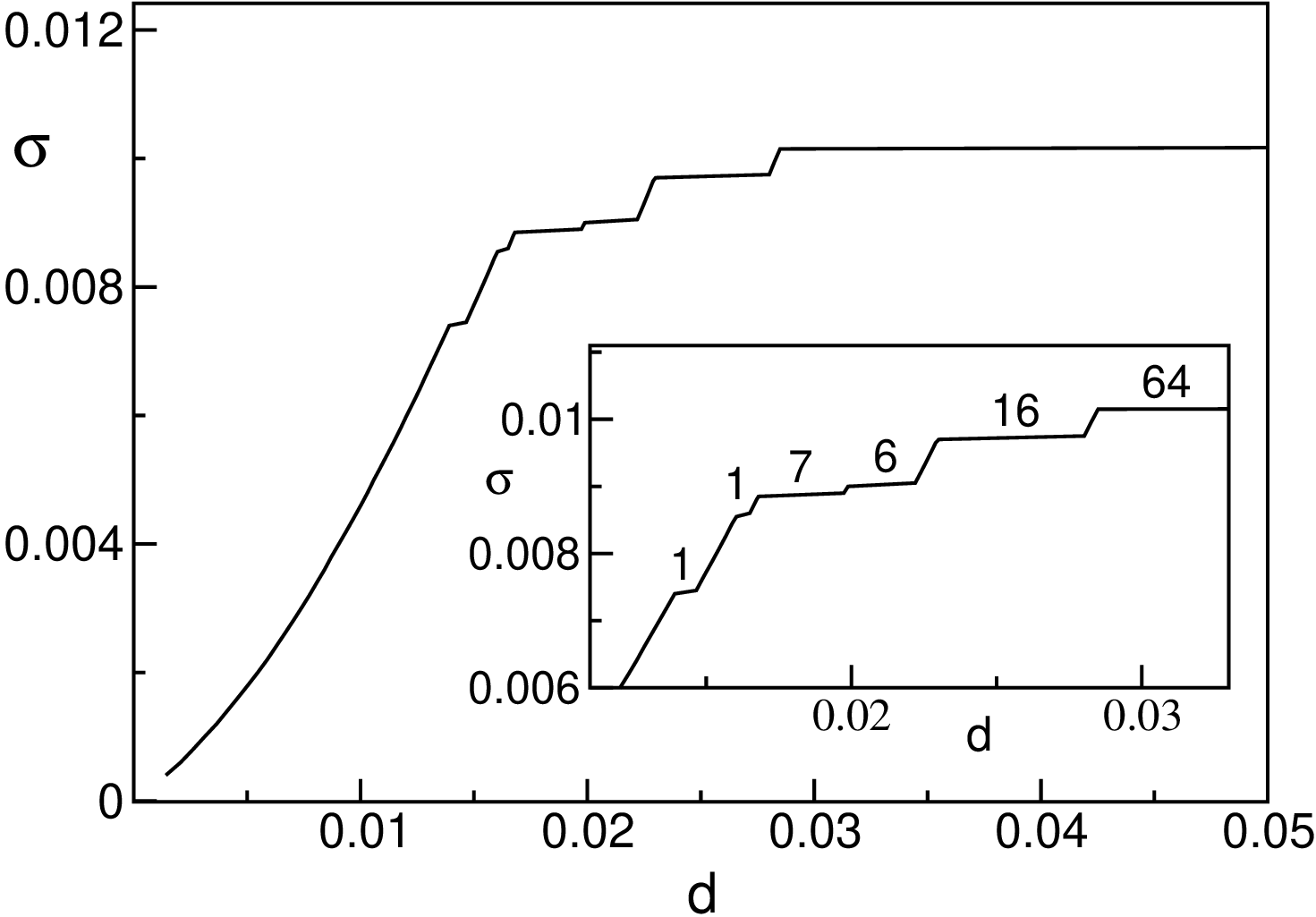}
\caption{The stress $\sigma$ as a function of the strain $d$, for one disorder
realization in a lattice with $40\times 42$ particles. Part of the curve is
zoomed in the inset so as to identify single avalanches (the numbers over the
horizontal steps represent the number of broken bonds in each avalanche).}
\label{load}
\end{figure}      
Since we are interested in investigating the fracture process in a nearly
static regime, a damping term $-\gamma\dot{\bf r}_i$ has been added 
to the forces acting on the $i$th particle. The last key element of the model is
disorder, which can be introduced by assuming that either the bond strengths
$k_{ij}$ or the thresholds $a^*_{ij}$ are distributed according to some law. 
Here, we assume $a_{ij}^*=a^*$ and a dichotomic distribution of bond strengths:
a fraction $c$ of them is set equal to 0 from the very beginning, while the rest
are set, without loss of generality, equal to 1. As pointed out in
Ref.~\cite{HRH89}, this choice is equivalent to a dichotomic distribution of
thresholds.

It is convenient to express all variables in adimensional units, since this
helps scaling out some parameters. In particular, the lattice spacing $a$ and
the particle masses can be both set equal to 1. This implies that the modulus
$\sigma$ of the applied force can be identified with the stress. A parameter
that cannot be scaled out is the threshold, here fixed equal to $a^*=1.04$.
This choice, besides being compatible with the linearization of the force field,
allows us reaching sizeable large-scale deviations from a purely crystalline
structure (we have been able to study lattice sizes up to $L=80$).

The equations of motion have been integrated by using a Runge-Kutta
algorithm with $\gamma=0.7$ and a time step equal to $0.3$. We have verified
that this choice guarantees the fastest convergence (in CPU-time units) to the
asymptotic state\cite{note1}. Finally, we have chosen to fix the fraction of
initially missing bonds equal to $c=0.3$, a value close to, but definitely
below, the rigidity-percolation threshold ($c_{r}=0.3398$). 

Numerical experiments consist in monitoring several observables while the 
stress is increased until a macroscopic failure occurs at $\sigma = \sigma_c$.
In order to ensure that the sampled configurations remain close to equilibrium
during the whole stretching process, (i) $\sigma$ is slowly increased, (ii)
additional relaxation loops are allowed when new bonds break. One of the
relevant observables is the strain
$d = \overline{|({\bf r}_i-{\bf r}_i^0)\cdot {\bf u}|}$, where the overline
denotes the average over all particles of the left and
right edges and ${\bf r}_i^0$ represents the equilibrium position.
As a typical example of the observed phenomenology, in Fig.~\ref{load}, we plot
the stress-strain curve for one realization of the disorder.
After an initially monotonous growth, a series of steps follows, each
corresponding to the ``simultaneous'' breaking of $s$ bonds. Since the same
amount of energy is released in each bond breaking, it is natural to identify
$s$ with $\varepsilon$ and interpret the phenomenon as the occurrence of an
avalanche.

\begin{figure}[tcb]
\centering
\includegraphics*[width=8 truecm, angle=0]{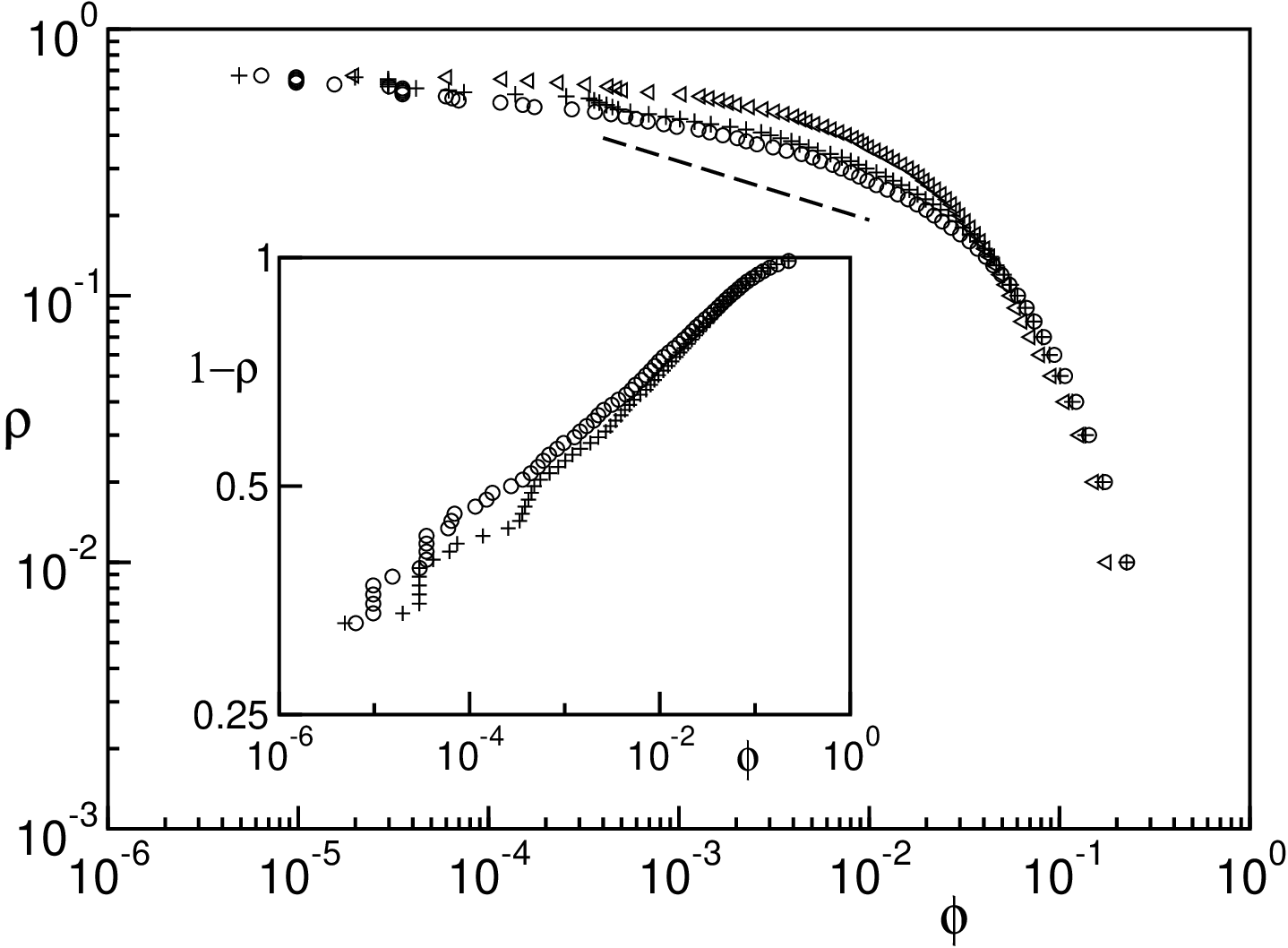}
\caption{Average cumulative fraction of broken bonds $\rho$ versus the
reduced stress $\phi$ for $L=20$ (triangles), 40 (plusses), and 80 (circles).
The dashed curve corresponds to the scaling behaviour observed in
Ref.~\cite{GGBC97}. In the inset, $1-\rho$ is plotted versus $\phi$ with the
same meaning of the symbols.}
\label{cumulative_energy}
\end{figure}      
Another relevant observable is the number $N(\sigma)$ of bonds broken at
stresses smaller or equal to $\sigma$, a quantity analogous to the cumulative
energy $E$ in Eq.~(\ref{energy}). Moreover, having the possibility to average
over many different realizations of the disorder (actually, we have studied
1000 realizations for each lattice size), it is also necessary to define a
meaningful way of performing ensemble averages. We have decided to consider the
fraction
\begin{equation}
\rho = \frac{N(\sigma)}{N(\sigma_c)}  \quad ,
\end{equation}
as the independent variable ($N(\sigma_c)$ being the total number of broken
bonds in each realization) and to average the stress values where $\rho$ is
attained. Moreover, analogously to Ref.~\cite{GGBC97}, where the reduced pressure
$P_r$ has been introduced, here we define the average reduced stress
\begin{equation}
\phi = \left \langle\frac{\sigma_{c}-\sigma}{\sigma_{c}} \right \rangle ,
\end{equation}
where $\langle \cdot \rangle$ denotes an average over disorder. Our simulations
indicate that, in the thermodynamic limit $L \to \infty$, the average becomes
irrelevant. Indeed, we have verified that $\sigma_c$ is a self-averaging
quantity, by observing that the normalized variance
\begin{equation}
\Delta_\sigma = \frac{\langle \sigma_c^2\rangle-\langle \sigma_c\rangle^2}
                  {\langle \sigma_c\rangle^2} ,
\end {equation}
decreases as $\Delta_\sigma \simeq L^{-2/3}$ with increasing $L$.

The resulting behavior for $L=20$, 40, and 80 is plotted in
Fig.~\ref{cumulative_energy}. The data for the $L=40$ and 80 nicely overlap,
indicating that finite-size effects are already negligible for $L \approx 40$.
In comparison with the experiment of Ref.~\cite{GGBC97}, the reduced stress here
covers a three-times wider range; this allows us ruling out the existence of
a power-law behaviour of $\rho$. On the other hand, if we restrict the analysis
to the range accessible in the experiment, we do not find relevant differences
with respect to the experiment itself, as indicated by the dashed line in
Fig.~\ref{cumulative_energy}. Next, one notices that $\rho$ remains strictly
smaller than 1 even close to $\phi =0$, suggesting that the 
fraction of bonds broken
in the macroscopic failure is finite. However, the most refined presentation
in the inset reveals that, though slowly, $\rho$ eventually converges to 1. 
Indeed, by fitting the small-$\phi$ region with $\rho=\rho_{0}-b\phi^{\eta}$,
we find that $\rho_0$ increases with $L$ and is equal already to 0.94 for
$L=80$. As for the $\eta$ value, our best estimate is $\eta = 0.12\pm 0.01$.
Although the fit is quite good, the smallness of $\eta$ leaves doubts about the
effective behaviour of $\rho$ in the vicinity of the macroscopic failure.

The dependence of $\rho$ on $\phi$ sheds light on the critical behaviour around
the onset of the macroscpic fracture. It is interesting to investigate also the
dependence of the total number of broken bonds $N(\sigma_c)$ on $L$. Here, we
have studied too a few sizes, to convincingly determine the scaling
behaviour; however, assuming that eventually $N(\sigma_c,L) = L^\alpha$, we
find that $\alpha \approx 1.5$, a value that is compatible with the simulations
of the same system with imposed strain \cite{PZ01}.

\begin{figure}[tcb]
\centering
\includegraphics*[width=8 truecm, angle=0]{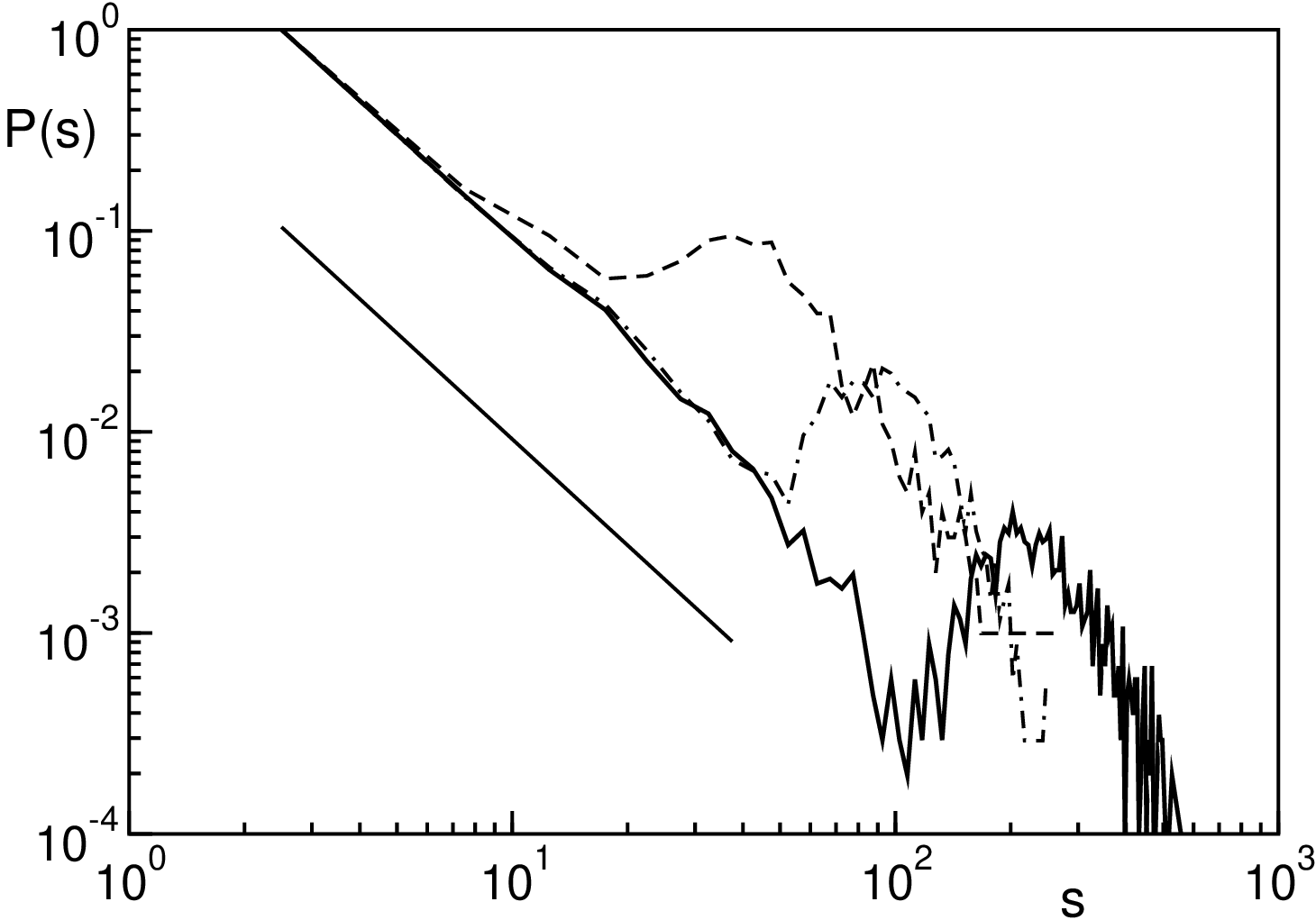}
\caption{Probability density $P(s)$ of avalanches preceding the macroscopic
failure, scaled to its maximum value. Dashed, dashed-dotted, and solid lines
refer to $L=20$, $40$, and $80$, respectively. The straight line, corresponding
to a decay $s^{-1.75}$ is the fit -- shifted for clarity -- of data for $L=80$}
\label{prob_density}
\end{figure}      
Another characterististic of fracture processes often studied both in
experiments and simulations is the distribution of microfractures (i.e.
avalanches). In the present context, this amounts to computing the number
$N_{s}$ of avalanches of size $s$; in order to increase the statistics, we
sum over all disorder realizations. The probability distributions for $L=20$,
40, and 80 are plotted in Fig.~\ref{prob_density}, where $N_s$ is scaled to its
maximum value ($P(s) = N_s/N_s^{max}$). All curves exhibit a seemingly
power-law decay followed by a peak at large $s$, which is clearly a finite-size
effect, since it corresponds to avalanche sizes that are comparable with the
lattice size. By fitting $P(s)$ in the meaningful $s$ range with $a s^{-\beta}$,
we find $\beta=1.75\pm0.06$, a value that lies inside the interval corresponding
to the various experimental measures, $[1.25,1.9]$, \cite{note2}.
\begin{figure}[tcb]
\centering
\includegraphics*[width=8 truecm, angle=0]{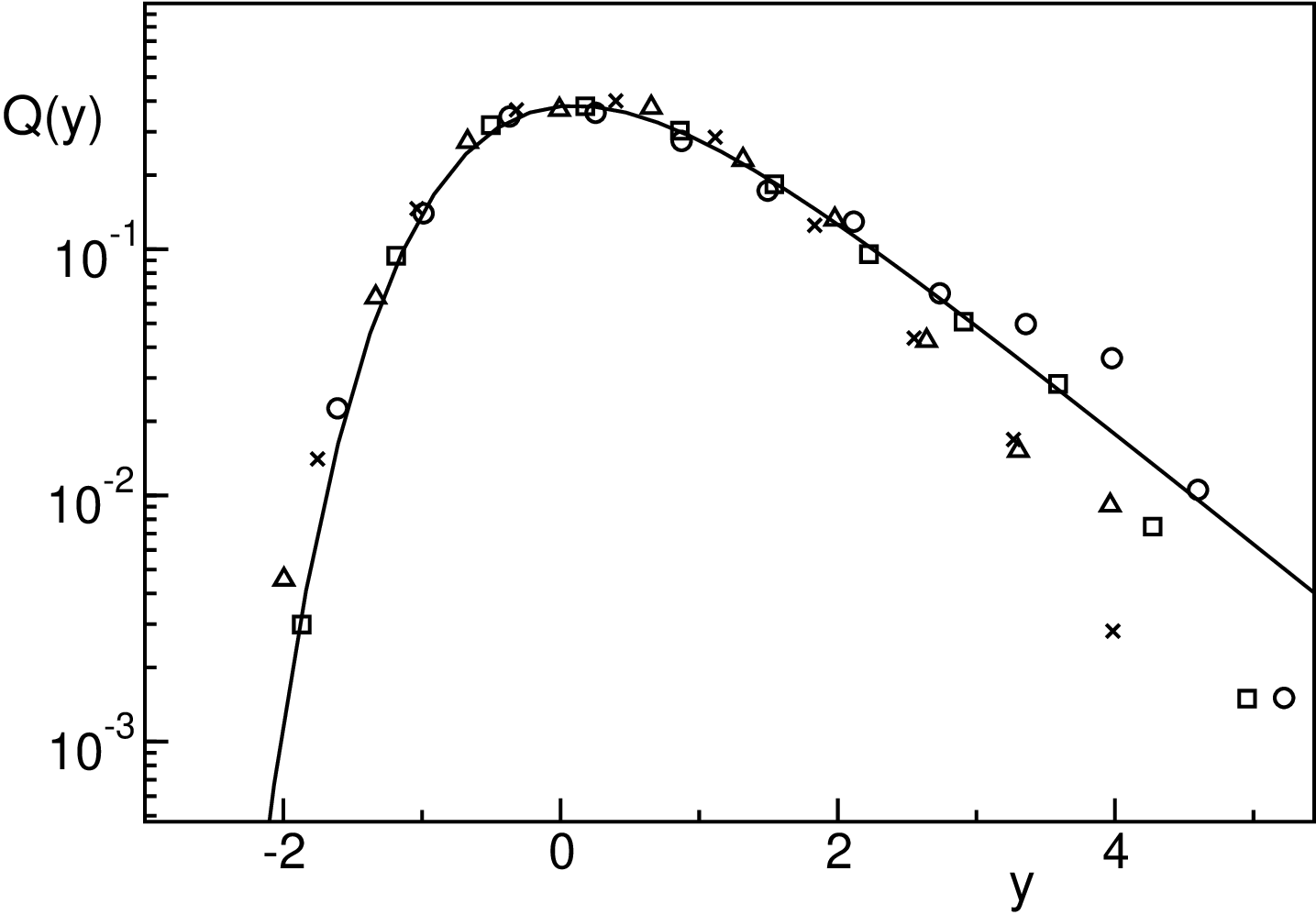}
\caption{The probability density $Q(y)$ of the final avalanche size $y$. The
size is shifted around the mean value and scaled to ensure a unit variance.
Crosses, triangles, squares and circles correspond to $L=10$, 20, 40, and 80.
The solid line corresponds to the Gumbel distribution.}
\label{gumbel}
\end{figure}      
Finally, we have investigated the size distribution of the macroscopic
avalanche. The results plotted in Fig.~\ref{gumbel} reveal increasing deviations
from a Gaussian behaviour, when $L$ is increased. The Gumbel distribution,
derived to describe extreme-value statistics (see the solid line) appears to
provide a more convincing description of the avalanche distribution.

Once the fracture process is recognized to be a non-equilibrium phase
transition, it becomes important to establish the order of the phenomenon.
From the existing literature, it is unclear whether we are in front of a second
order transition or the spinodal point of a first order transition
\cite{SA98,ZRSV99}. Such an uncertainty is mainly due to the fact that a
convincing order parameter has not yet been identified. On the basis of our
simulations and of the Lyon experiment, it seems reasonable to consider the
cumulative fraction of broken bonds as a proper order parameter. As a
consequence, since the contribution of 
the final avalanche is increasingly negligible in the large size limit,
we are led to conclude that the fracture is indeed a continuous transition.
However, it seems necessary to investigate yet larger systems in order to
convingly identify the asymptotic scaling behaviour of $\rho$ in the vicinity 
of the critical point.

While we are quite confident about the validity of our conclusions in this
specific model (since the same scenario arises when both the lattice
geometry and the direction of the applied stress is changed), we should add 
that a different behaviour has been found for rectangular distribution of
elastic constants. In such a case, although the numerical results are again 
compatible with the experimental ones (within the experimentally accessible 
range), $\rho(\phi)$ seems to exhibit a discontinuity at $\phi =0$. It thus 
not illogical to conjecture that the fracture may fall within (a few) 
different universality classes, depending on some microscopic details of 
the model that have still to be identifed.

This work has been partially funded by the FIRB-contract n. RBNE01CW3M\_001.



\begin{thebibliography}{10}
\bibitem{note1} The relaxation process becomes, however, increasingly slow, when
$L$ is increased and this problem should be seriously considered, if one were
interested in studying yet larger systems.
\end{thebibliography}
\end{document}